\documentclass[twoside,journey]{IEEEtran}
\usepackage{makecell}
\usepackage{hyperref}
\usepackage{array}
\usepackage{caption2}
\usepackage{graphicx,amssymb,amsmath}
\usepackage{multicol}
\usepackage[noadjust]{cite}
\usepackage{setspace}
\usepackage{subfigure}
\usepackage{graphicx}
\usepackage{float}
\usepackage{url}
\usepackage{stfloats}
\usepackage{amsthm,pifont}
\usepackage{flushend}
\usepackage{cases,subeqnarray}
\usepackage{bm,multirow,bigstrut}
\usepackage{amsmath, amsthm, amssymb}
\usepackage{textcomp}
\usepackage{latexsym,bm}
\usepackage{booktabs}
\usepackage{xcolor}
\usepackage{mathtools}
\usepackage{dsfont}
\usepackage{extarrows}
\usepackage{epsfig}
\usepackage{epsfig}
\usepackage{epstopdf}
\usepackage[noend]{algpseudocode}
\usepackage{algorithmicx,algorithm}
\theoremstyle{plain}

\theoremstyle{plain}
\newtheorem{rem}{Remark}
\newtheorem{them}{Theorem}

\usepackage{amsmath}

\newtheorem{lemma}{Lemma}
\IEEEoverridecommandlockouts

\begin{document}
	\title{Performance Analysis of Free-Space Information Sharing in Full-Duplex Semantic Communications}
	\author{Hongyang~Du, Jiacheng~Wang, Dusit~Niyato,~\IEEEmembership{Fellow,~IEEE}, Jiawen~Kang, Zehui Xiong, Dong~In~Kim,~\IEEEmembership{Fellow,~IEEE}, and Boon Hee Soong
		\thanks{H.~Du, J.~Wang, and D. Niyato are with the School of Computer Science and Engineering, Nanyang Technological University, Singapore (e-mail: hongyang001@e.ntu.edu.sg, jcwang\_cq@foxmail.com, dniyato@ntu.edu.sg)}
		\thanks{J. Kang is with the School of Automation, Guangdong University of Technology, China. (e-mail: kavinkang@gdut.edu.cn)}
		\thanks{Z. Xiong is with the Pillar of Information Systems Technology and Design, Singapore University of Technology and Design, Singapore (e-mail: zehui\_xiong@sutd.edu.sg)}
		\thanks{D. I. Kim is with the Department of Electrical and Computer Engineering, Sungkyunkwan University, South Korea (e-mail: dikim@skku.ac.kr)}
		\thanks{B. H. Soong is with the Department of Electrical and Electronic Engineering, Nanyang Technological University, Singapore (e-mail: ebhsoong@ntu.edu.sg)}
	}
	\maketitle
	\vspace{-1cm}
	\begin{abstract}
		In next-generation Internet services, such as Metaverse, the mixed reality (MR) technique plays a vital role. Yet the limited computing capacity of the user-side MR headset-mounted device (HMD) prevents its further application, especially in scenarios that require a lot of computation. One way out of this dilemma is to design an efficient information sharing scheme among users to replace the heavy and repetitive computation. In this paper, we propose a free-space information sharing mechanism based on full-duplex device-to-device (D2D) semantic communications. Specifically, the view images of MR users in the same real-world scenario may be analogous. Therefore, when one user (i.e., a device) completes some computation tasks, the user can send his own calculation results and the semantic features extracted from the user's own view image to nearby users (i.e., other devices). On this basis, other users can use the received semantic features to obtain the spatial matching of the computational results under their own view images without repeating the computation. Using generalized small-scale fading models, we analyze the key performance indicators of full-duplex D2D communications, including channel capacity and bit error probability, which directly affect the transmission of semantic information. Finally, the numerical analysis experiment proves the effectiveness of our proposed methods.
	\end{abstract}
	\IEEEpeerreviewmaketitle
	\section{Introduction}
\begin{figure}[t]
		\centering
		\includegraphics[width=0.3\textwidth]{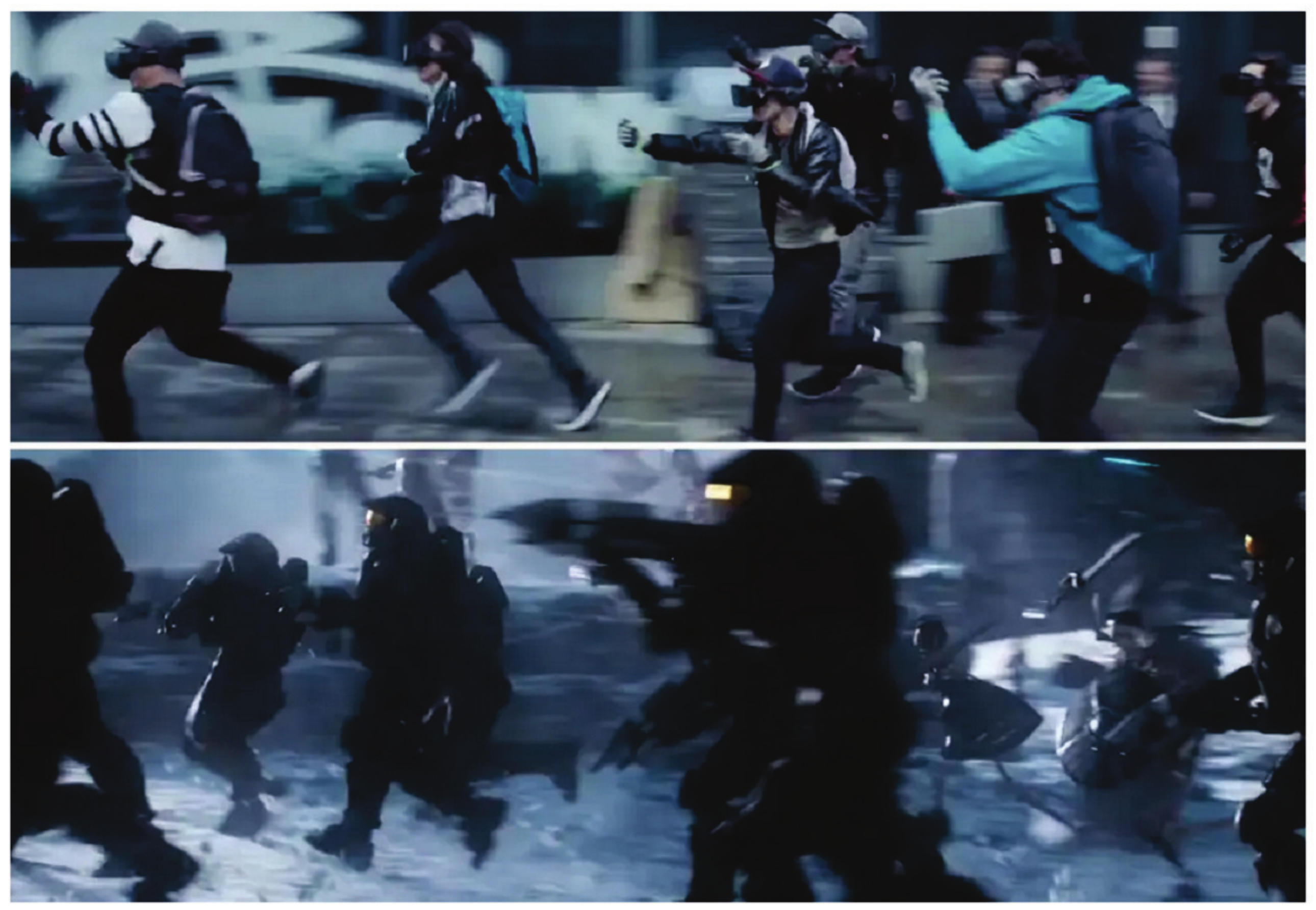}%
		\caption{In ``Ready Player One", MR-headset-wearing players run through the streets. However, they look like characters from the game ``Halo'' running into battle in the Oasis.}
		\label{ready}
	\end{figure}

	The development of computer science, wireless network, and immersive technologies, such as mixed reality (MR), have triggered the emergence of Metaverse, in which users can create their own avatars and experience different things in digital space. For instance, users wear MR equipment such as helmets and hold controllers to play a virtual battle game, just like the scene in the movie ``Ready Player One'' shown in the upper part of Fig. 1. During the game, the user's information in the physical world are collected and transmitted to Metaverse service provider (MSP). In Metaverse, the MSP leverages the obtained information to build a virtual gunfight scene, like the game ``Halo'' shown in the bottom half of Fig.~\ref{ready}, and display it to users through hardware devices.
	
	Although impressive, several difficulties need to be addressed before implementation. {\textit{One of the most representative and urgent problems to be solved is that the computational tasks of MR Head-Mounted Displays (HMDs) are over-heavy.}} A possible way out of this problem is to build an information sharing bridge among users to replace repeated calculations. For instance, in the virtual battle game shown in Fig.~\ref{ready}, the later player may follow in the footsteps of the previous one, indicating the later player would experience a similar scene that the previous one has already experienced. In such a case, we refer to the real-world scene that the user sees in the MR HMD, or that can be captured by the MR HMD camera, as the {\textit{view image}}. The detection of safe walk-able areas, i.e., {\textit{free-space information}}, in one user's MR HMD is based on its unique view image. Therefore, by sharing the computation results and the view image with other users, the users can determine the free-space in their own view images by comparing the received view image without repeating the computation redundantly. While feasible, transmitting such information among users requires a lot of transmission resources, especially for the high-definition view images.
	
	In response to the above mentioned a series of problems, in this paper, we propose a full-duplex semantic communications framework to achieve free space information sharing among users.  Concretely, the proposed semantic coding algorithm is employed to extract semantic features from the graph and then transmit it to other users in a full-duplex way. Using the semantic feature, users can obtain graphs that need to be rendered and displayed through lightweight algorithms, such as matching, instead of calculating independently. In this way, the computing resources consumption of each user is significantly reduced. Furthermore, the full duplex transmission of semantic information, which is extracted via calculation and has a smaller data amount than the original data, reduces user transmission resource consumption and facilitates implementation through MR devices worn by users.
	\begin{itemize}
		\item We propose a full-duplex D2D semantic communication based framework for synchronizing free-space information and semantic information among MR users to avoid each user independently computing the positions of virtual objects in the field of view. Here, the semantic information is the interesting points and corresponding descriptors of one user's view image.
		\item Within the proposed communication framework, we consider the impact of wireless fading and self-interference problems in D2D full-duplex communications on the transmission of semantic information and derive closed-form expressions of the bit error probability (BEP) and the channel capacity.
	\end{itemize}
	
	\section{System Model}\label{S3}
	In this section, we present the semantic-aware free-space information sharing mechanism and study a D2D full-duplex communications system.
	\subsection{Semantic-Aware Free-Space Information Sharing}
	\begin{figure}[t]
		\centering
		\includegraphics[width=0.3\textwidth]{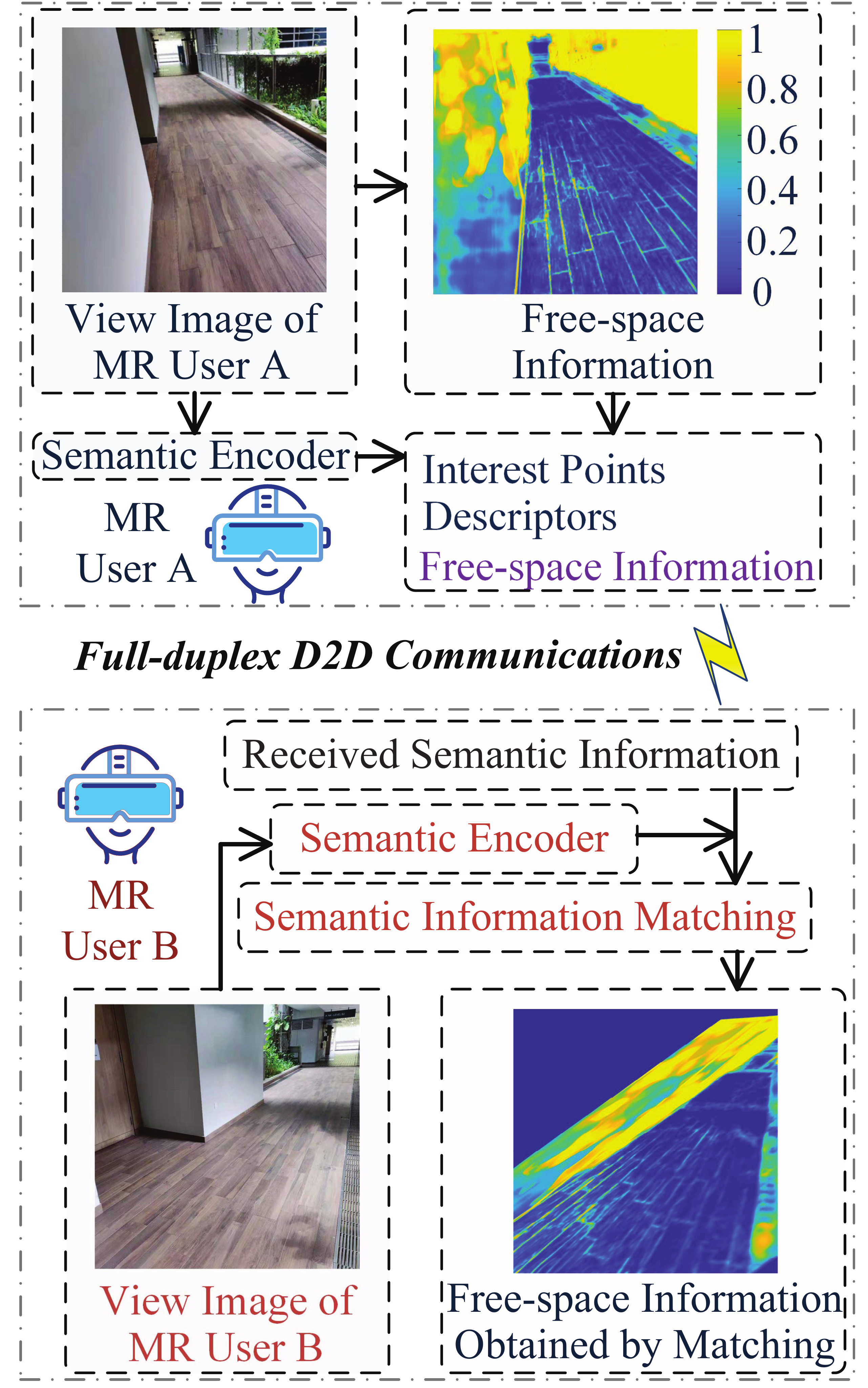}%
		\caption{The semantic-aware free-space information sharing mechanism proposed in this paper. Full-duplex wireless communication is performed between MR HMDs among users. The free-space information is computed by the state-of-art anomaly detection approach named JSR-Net~\cite{vojir2021road}}
		\label{model}
	\end{figure}
	When users use an MR technique to access virtual services, e.g., Metaverse, the MR HMDs need to perform various computing tasks. As shown in Fig.~\ref{ready}, HMD devices should calculate positions of virtual objects superimposed on the real world, and the planning of routes that users can walk safely. Since even the perspectives of users in the same scene may differ, each user needs to perform the above tasks independently. An alternative solution is to perform user view sharing. The shared users can obtain the sharer's computation results by view matching without repeating the computation independently. However, the sharing of original view images among MR HMDs consumes a lot of wireless transmission resources, i.e., transmit power and bandwidth. The reason is that the HMD resolution is high, resulting in a large amount of data for view images.

	To solve this problem and achieve effective information sharing, we apply the semantic communications in the D2D full-duplex communication system. A representative task could be {\textit{free-space information sharing}}. As shown in Fig.~\ref{model}, specifically, a user who identifies an area in their view image that is safe to walk in can share the detected free-space information with other users. Other users can then directly perform the matching to obtain safe walking area in their own view images. Let ${\bf{x}}_k$ denote the original view image of the $k^{\rm th}$ user. Here the semantic information includes the detected free-space (${\bf{i}}_k$), the interest points that reflect important structures in the visual field diagram (${{\bf{p}}_k}$), and the descriptors corresponding to interest points (${{\bf{d}}_k}$). Let $\mathcal{S}\left\{ {\cdot} \right\}$ denote the semantic encoder, we have $\mathcal{S}\left\{ {{{\bf{x}}_k}} \right\} = \left\{ {{{\bf{p}}_k},{{\bf{d}}_k},{{\bf{i}}_k}} \right\}$.
	The total data size of the semantic information, i.e., $\left\{ {{p_k},{d_k},{i_k}} \right\}$, will be much smaller than that of the original view image, i.e., $\left\{ {{x_k}} \right\}$. The detailed design of the semantic encoder is presented in Section~\ref{S4A}.

	
	\subsection{Full-Duplex Wireless Communications}
	Although semantic communication technologies can reduce the amount of data transmitted without affecting task completion, bandwidth resources are still strained when the number of users is large. Because of the limited bandwidth resources in the HMD device network, we consider that users are communicating with each other in the full-duplex mode.
	
	A critical issue in full-duplex communication is self-interference cancellation. Although various techniques have been proposed~\cite{tran2021special,perera2022sum}, it is still impossible to achieve perfect cancellation in edge devices such as MR HMDs. To investigate the effect of self-interference on system performance, we assume that the residual self-interference, i.e., ${{I}_k}$, is subject to the Gaussian distribution. This can be regarded as the worst-case assumption about the interference~\cite{han2020spectrum}. Specifically, the residual self-interference at $k^{\rm th}$ user and $j^{\rm th}$ user can be modeled as zero-mean complex Gaussian random variables with variance ${\upsilon _k}{P_k} \sigma_S^2$ and ${\upsilon _j}{P_k} \sigma_S^2$, respectively~\cite{han2020spectrum}, where ${P_k}$ is the transmit power of the $k^{\rm th}$ user and $\sigma_S^2$ denotes the variance due to unit power. The parameters ${\upsilon _k}$ and ${\upsilon _j}$ are constants that reflect the self-interference cancellation abilities of $k^{\rm th}$ user and $j^{\rm th}$ user, respectively~\cite{han2020spectrum}. 
	We consider that the $j^{\rm th}$ user is transmitting the semantic information to the $k^{\rm th}$ user. Let $s_j$ denote the transmitted symbol in $\mathcal{S}\left\{ {{{\bf{x}}_j}} \right\}$ The received signal at the $k^{\rm th}$ user can be expressed as
	{\small \begin{equation}
			{{y}_k} = \sqrt {{{P}_j}D_{jk}^{ - {\beta _k}}} {h_{jk}}{{s}_j} + \sum\limits_{i = 1}^N {\sqrt {{P_I}} } {g_{k}}{w_i} + {{I}_k} + {{n}_k},
	\end{equation}}\noindent
	where $ {n_k} \sim \mathcal{C}\mathcal{N}\left( {0,\sigma _N^2} \right) $, $ {I_k} \sim \mathcal{C}\mathcal{N}\left( 0 ,{\upsilon_k}{P_k} \sigma_S^2 \right) $, $h_{jk}$ is the small-scale fading channel, ${P_j}$ is the transmit power of the $j^{\rm th}$ user, $D_{jk}$ is the distance between the $k^{\rm th}$ user and $j^{\rm th}$ user, $\beta_{k}$ is the path loss exponent, $N$ is the number of interfere paths that is present at the $k_{\rm th}$ user-$j_{\rm th}$ user full-duplex communications pair, the average transmit power of each interfering signals is $P_{I}$, {{${w_{i}}$}} is the $i_{\rm th}$ interfering symbol, and $g_k$ is the small-scale fading of the interfering signal.
	
	We consider that the interference signals follow the Rayleigh distribution, i.e., $ {{{\left| {g_{k}} \right|}^2}} \sim {\rm Rayleigh} \left( {\eta _{k}}\right)$ . Note that the large scale fading of the interference signal is considered in the mean value of ${g_{k}}$. We then use the $\alpha-\mu$ distribution to model the small-scale fading, which is a general fading model that includes several important other distributions, such as the Weibull, One-Sided Gaussian, Rayleigh, and Nakagami. The probability density function (PDF) expression of a squared $\alpha - \mu$ random variable ${{\left| {{h_{jk}}} \right|}^2}$ are given by~\cite{yacoub2007alpha}:
	{\small \begin{equation}\label{PDFHJK}
			{f_{{\left| {{h_{jk}}} \right|}^2}}\left( x  \right) = \frac{{\alpha {x ^{\frac{{\alpha \mu }}{2} - 1}}}}{{2{\beta ^{\frac{{\alpha \mu }}{2} }}\Gamma \left( \mu  \right)}}\exp \left( { - {{\left( {\frac{x}{\beta }} \right)}^{\frac{\alpha }{2}}}} \right),
	\end{equation}}\noindent
where {$\Gamma\left(\cdot \right) $} is the gamma function \cite[eq. (8.310.1)]{gradshteyn2007}, {$ \beta  = \frac{{\bar \Upsilon \Gamma \left( \mu  \right)}}{{\Gamma \left( {\mu  + \frac{2}{\alpha }} \right)}} $}, and $\gamma\left(\cdot \right) $ is the incomplete gamma function \cite[eq. (8.35)]{gradshteyn2007}.
	The received signal-to-interference-plus-noise Ratio (SINR) of the $k^{\rm th}$ user can be expressed as
	{\small \begin{equation}\label{SINR}
			{\gamma _k} = \frac{{{P_j}D_{jk}^{ - {\beta _k}}{{\left| {{h_{jk}}} \right|}^2}}}{{{P_I}\sum\limits_{i = 1}^N {{{\left| {g_{k}} \right|}^2}}  + {\upsilon_k}{P_k}\sigma _S^2 + \sigma _N^2}}.
	\end{equation}}\noindent
	Simply, the SINR received by the $j^{\rm th}$ user can be obtained by interchanging $j$ and $k$ in \eqref{SINR}.

	\section{Semantic Encoding and Matching}\label{S4}
	In this section, we propose the semantic encoder design scheme according to SuperPoint~\cite{detone2018superpoint} and the semantic matching scheme according to SuperGlue~\cite{sarlin2020superglue}, using the received semantic information to perform view images matching.
	\subsection{Self-Supervised Semantic Encoding}\label{S4A}
	The definition of semantic information is typically task-relevant. For source messages in image modality, semantic information can be in the form of knowledge graphs~\cite{ji2021survey}, semantic segmentation results~\cite{ng2022stochastic}, or features~\cite{kang2022task}. In our proposed D2D free-space information sharing framework, the semantic information extracted from one user's view image needs to be used to help a second user determine the spatial correspondence between two view images.
	
	\begin{figure*}[t]
		\centering
		\includegraphics[width=0.76\textwidth]{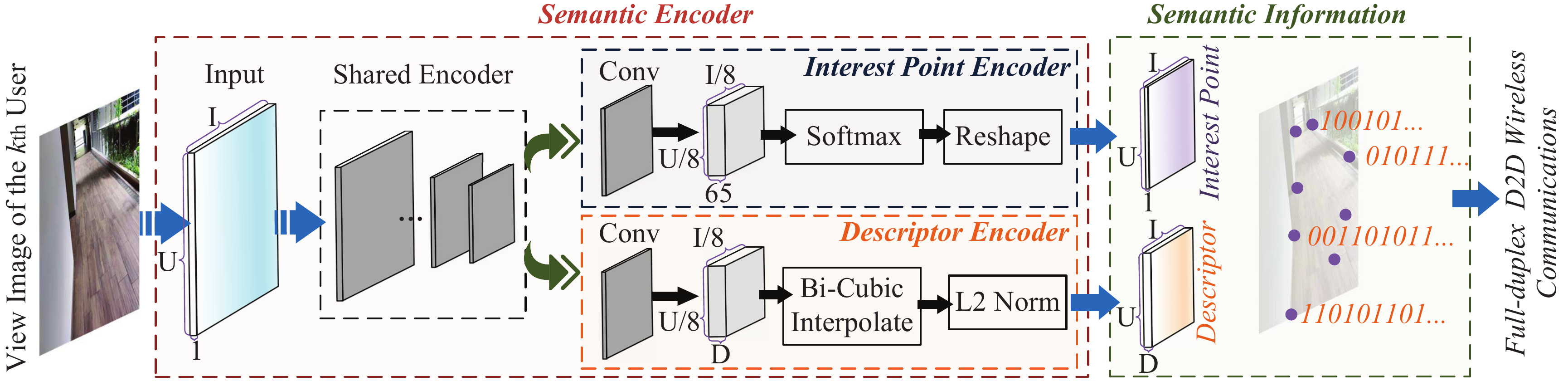}%
		\caption{The structure of the semantic encoder.}
		\label{SemanticEncoder}
	\end{figure*}
	Convolutional neural network (CNN) has been widely used in interest point detection and description. Here, we utilize the self-supervised SuperPoint~\cite{detone2018superpoint} architecture as the semantic encoder to produce interest point detection accompanied by fixed length descriptors in a single forward pass. Specifically, as shown in Fig.~\ref{SemanticEncoder}, the full-sized view image of the $k^{\rm th}$ user is sent to a shared CNN to reduce the dimensionality of the input view image. The output is then fed to two separate encoders to obtain the interest points and descriptors, respectively. After the network is trained, the required interest points and descriptors can be quickly obtained by inputting the user's view image. It has been shown that SuperPoint-based network architecture can support real-time interest points and descriptors extraction~\cite{detone2018superpoint}.
	
	The extracted interest points and descriptors are received and then used by other users via D2D full duplex wireless transmission. The performance metrics in wireless transmission are discussed in Section~\ref{S5}. We first discuss the semantic information matching problem after successful reception.
	
	\subsection{Semantic Information Matching}
	In the full-duplex system under our consideration, the semantic encoders discussed in the previous section should be deployed in each MR HMD, as shown in Fig.~\ref{model}. After other users receive the interest points and descriptors, semantic information extraction should also be performed on the user's own view image. Then, the semantic information matching is performed among the extracted interest points and descriptors. Here, we use SuperGlue~\cite{sarlin2020superglue} as a semantic matching network architecture. Note that a user may share its semantic information to multiple users in the form of D2D full-duplex wireless communication, each user is independent when matching semantic information. Therefore, in the following discussion we focus on the matching between the two view images.
	
	For a view image, we denote the set of keypoint positions $\bf{p}$ and corresponding descriptors $\bf{d}$ as $\left(\bf{p},\bf{d}\right) $. Here, the $\left(\bf{p},\bf{d}\right) $ obtained by the semantic encoder can be regarded as the local features. Specifically, the output of the interest point encoder consists of $x$ and $y$ image coordinates as well as a detection confidence $c$, ${\bf{p}}_i := \left( x; y; c\right)_i$. Visual descriptors ${\bf{d}}_i \in \mathbb{R}^D$ can be those extracted by a CNN like SuperPoint or traditional descriptors. 
	According to the SuperGlue~\cite{sarlin2020superglue}, for the integration into downstream tasks and better interpretability, each possible correspondence should have a confidence value. After training, the  SuperGlue model can be used in the user's MR HMDs to match semantic information, after receiving semantic features from the transmitter and extracting its own interest points and descriptors from the view image.
	
	\section{Performance Analysis And Resource Allocation}\label{S5}
	After obtaining the semantic information according to Section~\ref{S4A}, the user transmits the semantic information to other users in the form of D2D full-duplex communications. However, small-scale fading, interference and noise in wireless communication can affect channel capacity and BEP. To describe the above phenomenon, we first derive the PDF of the SINR in the D2D full-duplex communication system. The close-form channel capacity and BEP are also derived.
	\subsection{PDF expressions}
	To obtain the close-form expressions of the channel capacity and the BEP of the D2D full-duplex communication links, we first derive the PDF expression of the SINR.
	\begin{lemma}\label{lemma1}
		The close-form PDF expression of the SINR can be derived as
		{\small \begin{align}\label{PDFfinal}
				&{f_{{\gamma _k}}}\left( x \right) = \frac{{\alpha {x^{ - 1}}}}{{2\Gamma \left( N \right)\Gamma \left( \mu  \right)}}
				\notag\\&\times \!\!
				H_{1,0:1,1;1,1}^{0,0:0,1;1,1}\!\!\left(\!\!\!\!\!\! {\left. {\begin{array}{*{20}{c}}
							{{{\left(\! {\frac{{\varphi xD_{jk}^{{\beta _k}}}}{{\beta {P_j}}}} \!\right)}^{\frac{{ - \alpha }}{2}}}} \\ 
							{\frac{\varphi }{{{\eta _k}{P_I}}}} 
					\end{array}} \!\!\!\right|\!\!\!\!\!\begin{array}{*{20}{c}}
						{\left( {0;\! - \frac{\alpha }{2},1} \right)\!\!:\!\!\left(\! {1 - \mu ,1} \right)\!;\!\left( {1,1} \right)} \\ 
						{ - :\left( {1,\frac{\alpha }{2}} \right);\left( {N,1} \right)} 
				\end{array}} \!\!\!\!\!\right)\!,
		\end{align}}\noindent
		where $H_{ \cdot  \cdot }^{ \cdot  \cdot }\left( { \cdot \left|  \cdot  \right.} \right)$ is the Multivariate Fox's $H$-function \cite[eq. (A-1)]{mathai2009h}.
		\begin{IEEEproof}
Because $ {{{\left| {g_{k}} \right|}^2}} \sim {\rm Rayleigh} \left( {\eta_{k}}\right)$ and the sum of $N$ i.i.d. Rayleigh-fading signals have a Nakagami-$m$ distributed signal amplitude with $m = N$, the PDF of $ U \triangleq {\sum\limits_{i = 1}^N {{{\left| {g_{k}} \right|}^2}} } $ \cite{nakagami1960m} can be written as $f_{U}\left( x \right) = \frac{{{x^{N - 1}}}}{{\eta _{k}^N\Gamma \left( N \right)}}\exp \left( { - \frac{x}{{{\eta _{k}}}}} \right)$, where $ {{\eta_{k}}}={{{\mathbb E}\left[ g_{k}^2 \right]}}$, ${\mathbb E}\left[ \cdot \right]$ denotes expectation, and $\gamma\left(\cdot,\cdot\right)$ is the incomplete Gamma function defined in \cite[eq. (8.350.1)]{gradshteyn2007}.
Let $ V \triangleq {P_I}U + {\upsilon_k}{P_k}\sigma _S^2 + \sigma _N^2 $ and $\varphi \triangleq {{\upsilon_k}{P_k}\sigma _S^2 + \sigma _N^2} $. Then, the PDF of $V$ can be derived as ${f_V}\left( y \right) = \frac{1}{{{P_I}}}{f_U}\left( {\frac{{y - \varphi}}{{{P_I}}}} \right)$. Substituting $f_{U}\left( x \right)$ into ${f_V}\left( y \right)$, we have
{\small \begin{equation}
		{f_V}\left( y \right) = \frac{{{{\left( {y - \varphi} \right)}^{N - 1}}}}{{{{\left( {{P_I}{\eta _k}} \right)}^N}\Gamma \left( N \right)}}\exp \left( { - \frac{{y - \varphi}}{{{\eta _k}{P_I}}}} \right).
\end{equation}}\noindent
Let $ W \triangleq {P_j}D_{jk}^{ - {\beta _k}}{\left| {{h_{jk}}} \right|^2} $. We can obtain that ${f_W}\left( w \right) = \frac{1}{{{P_j}D_{jk}^{ - {\beta _k}}}}{f_{{{\left| {{h_{jk}}} \right|}^2}}}\left( {\frac{w}{{{P_j}D_{jk}^{ - {\beta _k}}}}} \right)$. By substituting \eqref{PDFHJK} into ${f_W}\left( w \right)$,the PDF of $W$ can be derived as
{\small \begin{equation}
		{f_W}\left( w \right) = \frac{{\alpha {{\left( w \right)}^{\frac{{\alpha \mu }}{2} - 1}}}}{{2{{\left( {{P_j}D_{jk}^{ - {\beta _k}}\beta } \right)}^{\frac{{\alpha \mu }}{2}}}\Gamma \left( \mu  \right)}}\exp \left( { - {{\left( {\frac{w}{{\beta {P_j}D_{jk}^{ - {\beta _k}}}}} \right)}^{\frac{\alpha }{2}}}} \right).
\end{equation}}\noindent
Because $ {\gamma _k} = \frac{W}{V} $, we can derive the PDF of ${\gamma _k}$ as ${f_{{\gamma _k}}}\left( x \right) = \int_0^\infty  {y{f_W}\left( {xy} \right){f_V}\left( y \right){\rm{d}}y}$. Using ${f_W}\left( w \right)$ and ${f_V}\left( y \right)$, we have
{\small \begin{equation}\label{pdfd}
		{f_{{\gamma _k}}}\left( x \right) = \frac{{\alpha {x^{\frac{{\alpha \mu }}{2} - 1}}}}{{2{{\left( {{P_j}D_{jk}^{ - {\beta _k}}\beta } \right)}^{\frac{{\alpha \mu }}{2}}}{{\left( {{P_I}{\eta _k}} \right)}^N}\Gamma \left( N \right)\Gamma \left( \mu  \right)}}{I_1},
\end{equation}}\noindent
where
{\small \begin{align}\label{i1}
		{I_1} = \!\! \int_{\varphi }^\infty  \!\!{{y^{\frac{{\alpha \mu }}{2}}}{{\left( {y\! - \!\varphi} \right)}^{N - 1}}} \exp\!\left(\! { \frac{{ \varphi- y}}{{{\eta _k}{P_I}}}} \!\right)
\!\exp\!\left(\! { - {{\left(\! {\frac{{xy}}{{\beta {P_j}D_{jk}^{ - {\beta _k}}}}} \!\right)}^{\frac{\alpha }{2}}}}\! \right)\!{\rm{d}}y.
\end{align}}\noindent
To solve for the integral in ${I_1}$, we map the exponential function to the complex domain \cite[eq. (01.03.07.0001.01)]{web}. By changing the order of integration, with the help of~\cite[eq. (3.191.2)]{gradshteyn2007}, $I_1$ can be solved.
By substituting \eqref{i1} and \eqref{i2} into \eqref{pdfd} and using \cite[eq. (8.384.1)]{gradshteyn2007} and \cite[eq. (A-1)]{mathai2009h}, the PDF of ${\gamma _k}$ can be expressed in form of the Multivariate Fox's $H$-function as \eqref{PDFfinal} to complete the proof.
		\end{IEEEproof}
	\end{lemma}
	\subsection{Channel Capacity in Full-duplex Communications}
	The ergodic channel capacity (or Shannon capacity) for the transmitter, which is known to be the maximum data rate that the channel can support (per Hz), is defined as $C_{\rm k} = \int_0^\infty  {{{\log }_2}} (1 + x ){f_{\gamma_k} }(x ){\rm{d}}x$. Then, with the help of Lemma~\ref{lemma1}, the channel capacity can be obtained in the following theorem.
	\begin{them}\label{Theorem1}
		The channel capacity of the $k_{\rm th}$ user, i.e., ${C_{\rm{k}}}$, can be derived in close-form as
		{\small \begin{align} \label{ccfinal}
				&{C_{\rm{k}}} = \frac{{{\Gamma ^{ - 1}}\left( \mu  \right)\alpha }}{{\ln 4\Gamma \left( N \right)}}
				\notag\\&\times\!\!
				H_{1,0:3,3;1,1}^{0,0:1,3;1,1}\!\!\!\left(\!\!\!\!\!\! {\left. {\begin{array}{*{20}{c}}
							{{{\left(\!\! {\frac{{\beta {P_j}}}{{\varphi D_{jk}^{{\beta _k}}}}} \!\!\right)}^{\frac{\alpha }{2}}}} \\ 
							{\frac{\varphi }{{{\eta _k}{P_I}}}} 
					\end{array}} \!\!\!\!\right|\!\!\!\!\!\begin{array}{*{20}{c}}
						{\left( {0;\!- \frac{\alpha }{2},\!1} \right)\!\!:\!\!\left(\!{1 \!-\! \mu ,\!1} \right)\!\!\left( {1,\frac{\alpha }{2}} \right)\!\!\left( {1,\frac{\alpha }{2}} \right)\!;\!\left( {1,\!1} \right)} \\ 
						{ - :\left( {1,\frac{\alpha }{2}} \right)\left( {0,\frac{\alpha }{2}} \right)\left( {1,\frac{\alpha }{2}} \right);\left( {N,1} \right)} 
				\end{array}} \!\!\!\!\!\right)\!.
		\end{align}}\noindent
		\begin{IEEEproof}
According to the definition of channel capacity, we can express the channel capacity as
{\small \begin{align}\label{ccguo}
		{C_{\rm{k}}} =& \frac{{{\Gamma ^{ - 1}}\left( \mu  \right)\alpha }}{{2\Gamma \left( N \right)}}{\left( {\frac{{\text{1}}}{{{\text{2}}\pi i}}} \right)^2}\int_{{\mathcal{L}_1}} {\int_{{\mathcal{L}_2}} {\frac{{\Gamma \left( {{t_1} + \mu } \right)\Gamma \left( {{t_2}} \right)\Gamma \left( {N - {t_2}} \right)}}{{\Gamma \left( {\frac{\alpha }{2}{t_1} - {t_2}} \right)\Gamma \left( {\frac{\alpha }{2}{t_1}} \right)}}} } 
		\notag\\&\times
		I_4{\left( {\frac{\varphi }{{\beta {P_j}D_{jk}^{ - {\beta _k}}}}} \right)^{\frac{{ - \alpha }}{2}{t_1}}}{\left( {\frac{\varphi }{{{\eta _k}{P_I}}}} \right)^{{t_2}}}{\rm{d}}{t_1}{\rm{d}}{t_2}
\end{align}}\noindent
where ${I_4} = \int_0^\infty  {{y^{\frac{{ - \alpha }}{2}{t_1} - 1}}\log_2 \left( {1 + y} \right){\rm{d}}y}$. Using \cite[eq. (2.6.9.21)]{Prudnikov1986Integrals}, \cite[eq. (8.334.3)]{gradshteyn2007} and \cite[eq. (A-1)]{mathai2009h}, we can derive the channel capacity as \eqref{ccfinal} to complete the proof.
		\end{IEEEproof}
	\end{them}
	
	\subsection{Bit Error Probability in Full-duplex Communications}
	The BEP for the $k^{\rm th}$ user under a variety of modulation formats is given by~\cite{tse2005fundamentals} as $E_{\rm k} = \int_0^\infty  {\frac{{\Gamma \! \left( {{\tau _2},{\tau _1}\gamma } \right)}}{{2\Gamma \! \left( {{\tau _2}} \right)}}{f_{\gamma_k}}\left( \gamma  \right)d\gamma }$,
	where $\Gamma \left( { \cdot , \cdot } \right)$ is the upper incomplete Gamma function \cite[eq. (8.350.2)]{gradshteyn2007}, $ {{{\Gamma \left( {{\tau _2},{\tau _1}\gamma } \right)}}/{{2\Gamma \left( {{\tau _2}} \right)}}} $ is the conditional BEP ${\tau _1} $ and ${\tau _2}$ are modulation-specific parameters that represents various modulation/detection combinations. Specifically, $\left\{ {{\tau _1} = 1,{\tau _2} = 0.5} \right\}$ is antipodal coherent binary phase-shift keying (BPSK), $\left\{ {{\tau _1} = 0.5,{\tau _2} = 0.5} \right\}$ is orthogonal coherent binary frequency-shift keying (BFSK), $\left\{ {{\tau _1} = 0.5,{\tau _2} = 1} \right\}$ is orthogonal non-coherent BFSK, and $\left\{ {{\tau _1} = 1,{\tau _2} = 1} \right\}$ denote antipodal differentially coherent BPSK (DPSK).
	
	\begin{them}\label{Theorem2}
		The close-form expression of the BEP of the $k_{\rm th}$ user, i.e., $E_k$, is derived as
		{\small \begin{align}\label{berfinal}
				&	{E_k} = \frac{{\alpha {\tau _1}^{\frac{\alpha }{2}{t_1}}{\Gamma ^{ - 1}}\left( {{\tau _2}} \right)}}{{4\Gamma \left( N \right)\Gamma \left( \mu  \right)}}
				\notag\\&\times\!\!
				H_{1,1:2,2;1,1}^{0,1:1,1;1,1}\!\!\left(\!\!\!\!\!\! {\left. {\begin{array}{*{20}{c}}
							{{{\left(\! {\frac{{\beta {P_j}}}{{\varphi D_{jk}^{{\beta _k}}}}} \!\right)}^{\frac{\alpha }{2}}}} \\ 
							{\frac{\varphi }{{{\eta _k}{P_I}}}} 
					\end{array}} \!\!\!\!\right|\!\!\!\!\begin{array}{*{20}{c}}
						{\left( {1; - \frac{\alpha }{2},1} \right)\!\!:\!\left( {1\! - \!\mu ,1} \right)\left( {1,\frac{\alpha }{2}} \right)\!;\!\left( {1,1} \right)} \\ 
						{\left( {1;\frac{\alpha }{2}, - 1} \right)\!:\!\left( {0,\frac{\alpha }{2}} \right)\left( {1,\frac{\alpha }{2}} \right);\left( {N,1} \right)} 
				\end{array}} \!\!\!\!\!\right).
		\end{align}}\noindent
		\begin{IEEEproof}
	With the help of the definition of Gamma function \cite[eq. (8.350)]{gradshteyn2007}, we can re-write $E_k$ as $E_k = \frac{{{\tau _1}^{{\tau _2}}}}{{2\Gamma \left( {{\tau _2}} \right)}}\int_0^\infty  {{x^{{\tau _2} - 1}}{e^{ - {\tau _1}x}}{F_{{\gamma _{k}}}}\left( x \right){\rm{d}}x}$. Then, we obtain
{\small \begin{align}\label{faeg3}
		&	{E_k} = \int_{{\mathcal{L}_1}} {\int_{{\mathcal{L}_2}} {\frac{{\Gamma \left( {{t_1} + \mu } \right)\Gamma \left( { - \frac{\alpha }{2}{t_1}} \right)\Gamma \left( {{t_2}} \right)\Gamma \left( {N - {t_2}} \right)}}{{\Gamma \left( {\frac{\alpha }{2}{t_1} - {t_2}} \right)\Gamma \left( {\frac{\alpha }{2}{t_1}} \right)\Gamma \left( {{\text{1}} - \frac{\alpha }{2}{t_1}} \right)}}} } {I_5}
		\notag\\&\times\!\!\!
		\frac{{\alpha {\tau _1}^{{\tau _2}}{\Gamma ^{ - 1}}\left( {{\tau _2}} \right)}}{{4\Gamma\!\left( N \right)\Gamma\!\left( \mu  \right)}}{\left(\! {\frac{{\text{1}}}{{{\text{2}}\pi i}}} \!\right)^2}{\left(\! {\frac{{D_{jk}^{{\beta _k}}\varphi }}{{\beta {P_j}}}} \!\right)^{\frac{{ - \alpha }}{2}{t_1}}}{\left(\! {\frac{\varphi }{{{\eta _k}{P_I}}}} \!\right)^{{t_2}}}{\rm{d}}{t_1}{\rm{d}}{t_2}
\end{align}}\noindent
where ${I_5} = \int_0^\infty  {{x^{\frac{{ - \alpha }}{2}{t_1} + {\tau _2} - 1}}{e^{ - {\tau _1}x}}{\text{d}}x}$. With the help of~\cite[eq. (3.351.3)]{gradshteyn2007} and~\cite[eq. (8.339.1)]{gradshteyn2007}, $I_D$ can be solved. Substituting $I_5$ into \eqref{faeg3} and using~\cite[eq. (A-1)]{mathai2009h}, we can derive \ref{berfinal} to complete the proof.
		\end{IEEEproof}
	\end{them}
	\begin{rem}
		The above derived performance indicator expressions contain the important channel parameters in a D2D full-duplex wireless communications system. Specifically, parameters that can respond to large-scale fading, small-scale fading, self-interference in duplex communications, and co-channel interference at arbitrary multipath are all included in \eqref{ccfinal} and \eqref{berfinal}. Detailed discussion is given in Section~\ref{per}. Furthermore, the generalization of the small-scale fading model, i.e., $\alpha$-$\mu$ fading channel model, allows us to use the derived performance metrics in the analysis of many systems with various kinds of fading conditions.
	\end{rem}

	\section{Numerical Results}\label{S6}
	In this paper, we propose a device-to-device semantic information sharing mechanism, where the semantic information is the interest points and descriptors of one user's view image.
	\subsection{Effectiveness of semantic communication framework}
\begin{figure*}[t]
		\centering
		\includegraphics[width=0.9\textwidth]{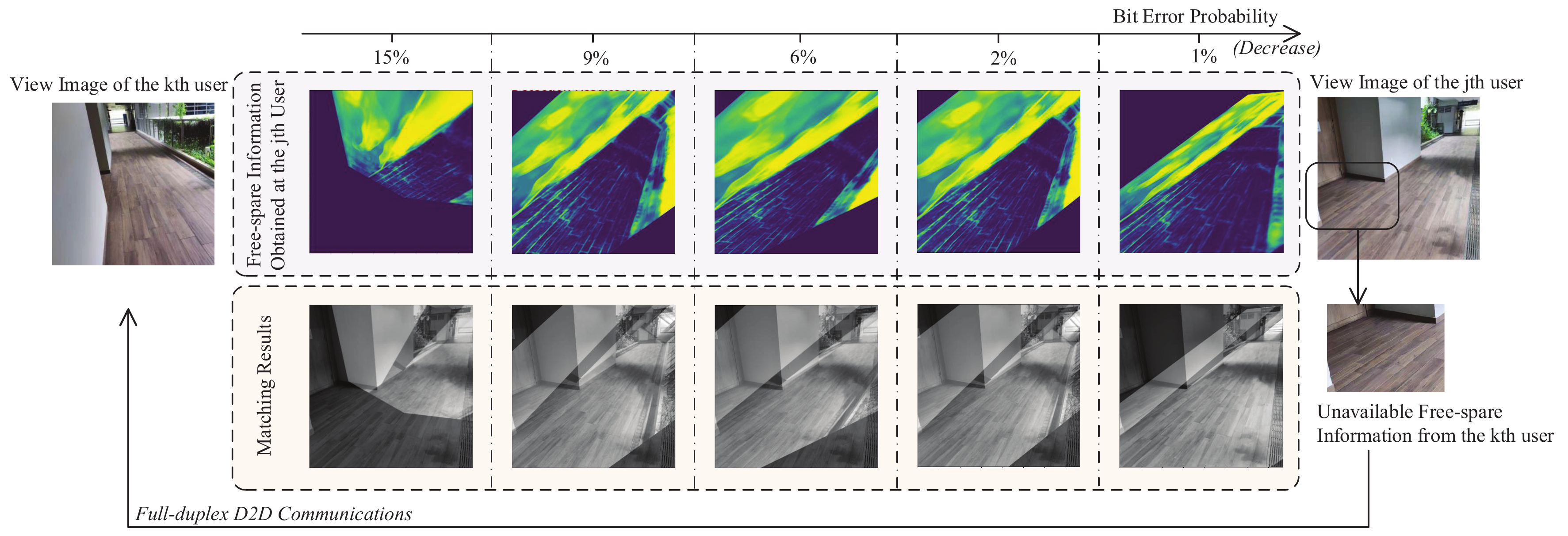}%
		\vspace{-0.2cm}
		\caption{The effect of our proposed free-space sharing mechanism with different wireless transmission BEP.}
		\label{match}
	\end{figure*}

	First, we verify the effectiveness of the proposed D2D full-duplex semantic communications framework, and the results are shown in Fig.~\ref{match}. Firstly, according to the interest points, descriptors, and free-space information transmitted by the ${k^{\rm th}}$ user, the ${j^{\rm th}}$ user can effectively obtained his own free-space information, as shown by the images in the first row. Compared with the conventional approach (i.e., the ${j^{\rm th}}$ user computes the free-space information independently), this method consumes fewer resources since the ${j^{\rm th}}$ user only needs to calculate the spatial correspondence information between view images to obtain his own free-space information. Secondly, the results demonstrate that as the BEP decreases, the free-space information calculated by the ${j^{\rm th}}$ user is more accurate, which finally generates a more accurate matching result, as presented by the images in the second row. Note that the values of BEP can be calculated by using the environmental parameters and \eqref{berfinal}. This inspires us that the accuracy of semantic information during wireless transmission needs to be guaranteed in the free-space information sharing mechanism.
	
	It should be noted that due to the limitation of the viewing angle, some details, such as the corner marked by the black box in the ${j^{\rm th}}$ user's view, cannot be obtained using the semantic information extracted from the view image of the ${k^{\rm th}}$ user. However, this part of the information may be obtained through semantic information from users who have other different perspectives. Meanwhile, because full-duplex communication is taken into account here, the ${j^{\rm th}}$ user would feed back the missing information to the ${k^{\rm th}}$ user and other users, allowing the ${k^{\rm th}}$ user to obtain more free-space information, even if there is a blind spot. Such real-time information updating mechanism is also an advantage of our proposed framework.
	\begin{figure*}
		\setcaptionwidth{2.25in}
		\setlength{\abovecaptionskip}{-2pt}
		\setlength{\belowcaptionskip}{-1pt}
		\centering
		\begin{minipage}[t]{0.331\linewidth} 
			\centering
			\includegraphics[width=0.9\textwidth]{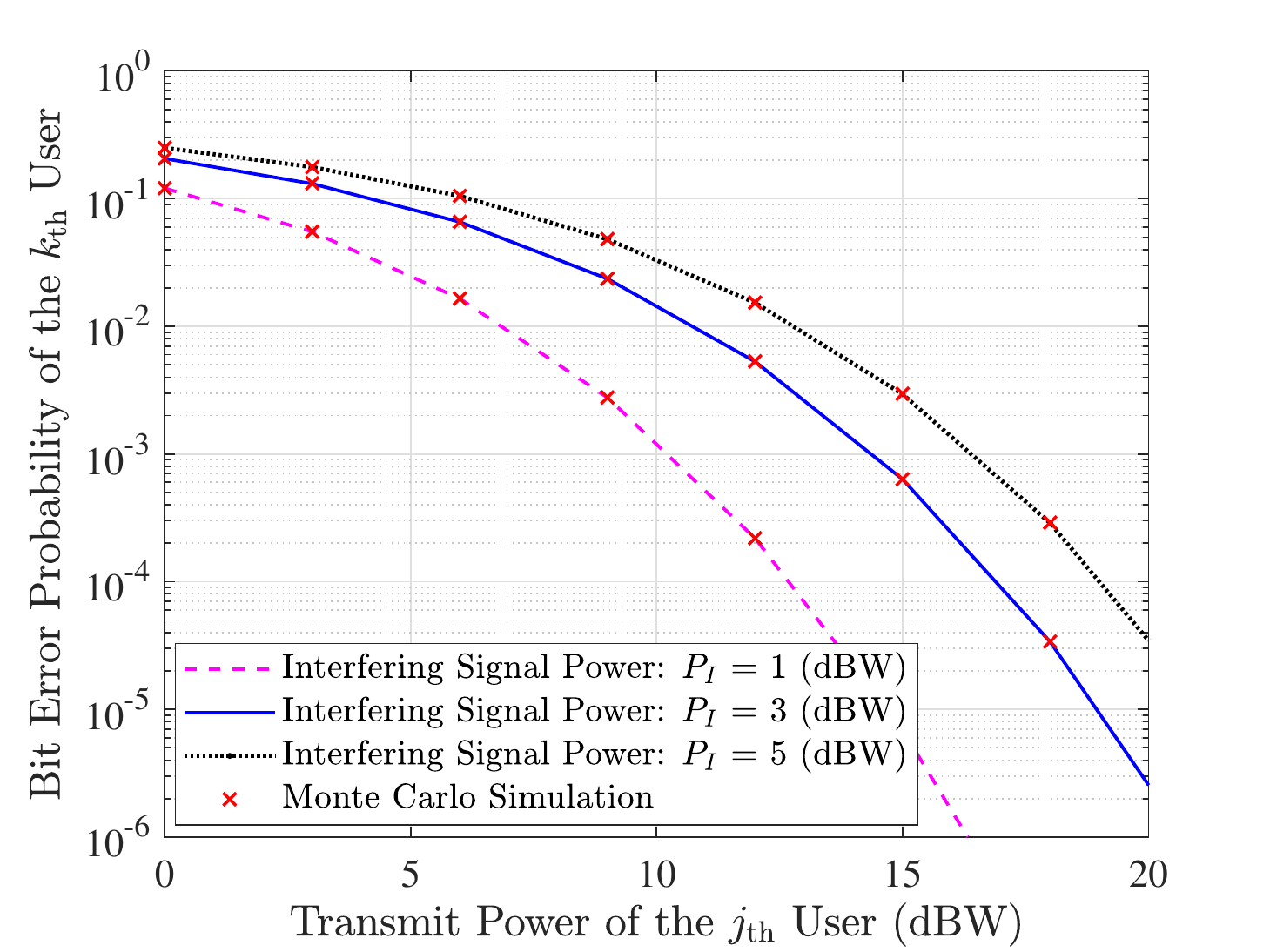}%
			\vspace{0.05cm}
			\caption{The BEP of the $k^{\rm th}$ user versus the transmit power of the $j^{\rm th}$ user under different interfering power, with $D_{jk} = 5$ m, $\beta_{k} = 2$, $\alpha = 4$, $\mu = 5$, $\sigma _N^2 = \sigma _S^2 = 0.1$, $P_k = 20$ ${\rm dBW}$, $P_I = 1$ ${\rm dBW}$, ${\eta _{k}} = 0.2$, ${\upsilon _k}=0.2$, and $N = 2$.}
			\label{BER1}
		\end{minipage}%
		\begin{minipage}[t]{0.331\linewidth}
			\centering
			\includegraphics[width=0.9\textwidth]{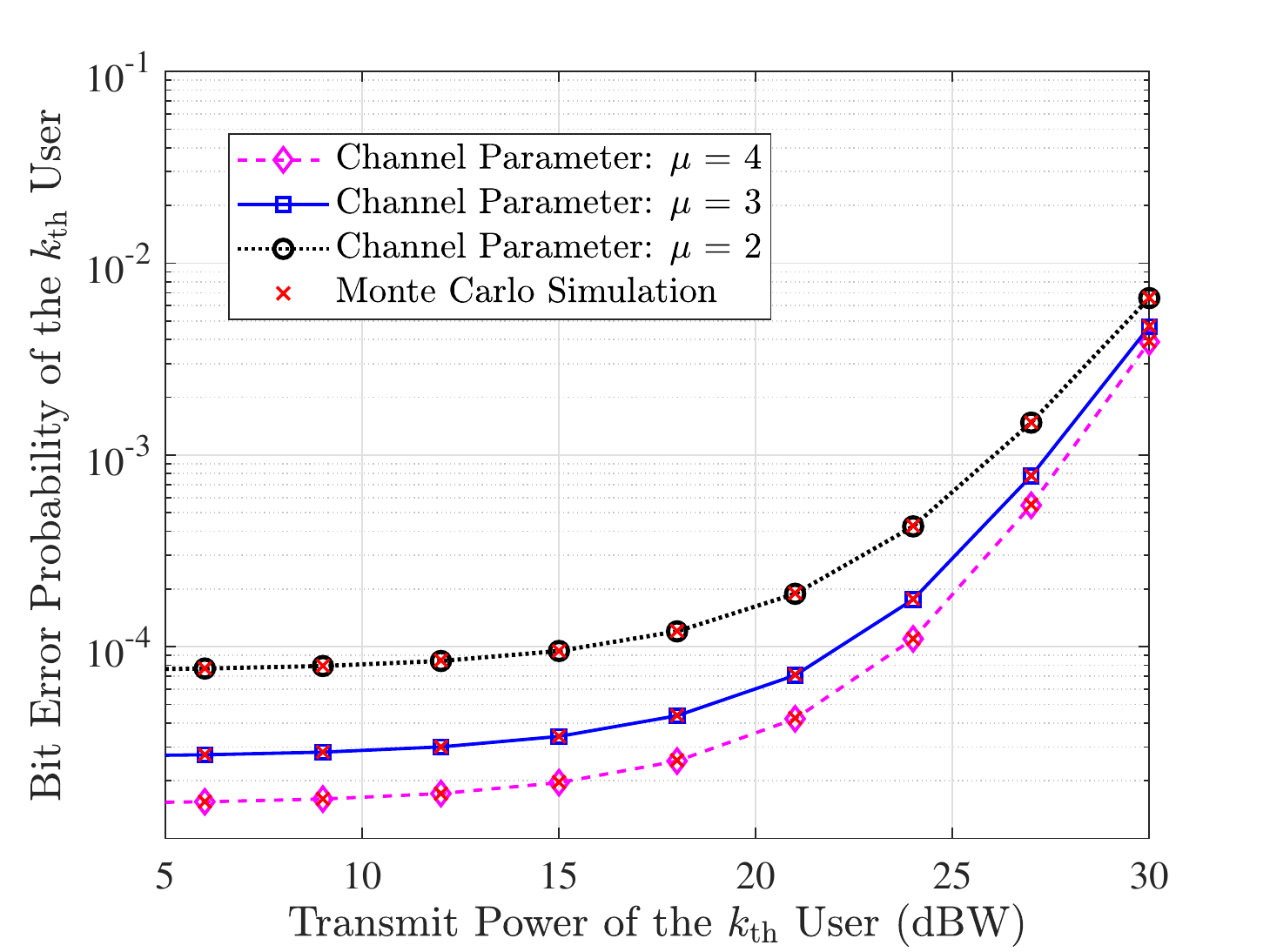}%
			\vspace{0.05cm}
			\caption{The BEP of the $k^{\rm th}$ user versus the transmit power of the $k^{\rm th}$ user under different channel parameters, with $D_{jk} = 10$ m, $\beta_{k} = 2$, $\alpha = 4$, $\mu = 4$, $\sigma _N^2 = \sigma _S^2 = 0.1$, $P_j = 20$ ${\rm dBW}$, $P_I = 1$ ${\rm dBW}$, ${\eta _{k}} = 0.2$, ${\upsilon _k}=0.2$, and $N = 2$.}
			\label{BER2}
		\end{minipage}
		\begin{minipage}[t]{0.331\linewidth}
			\centering
			\includegraphics[width=0.9\textwidth]{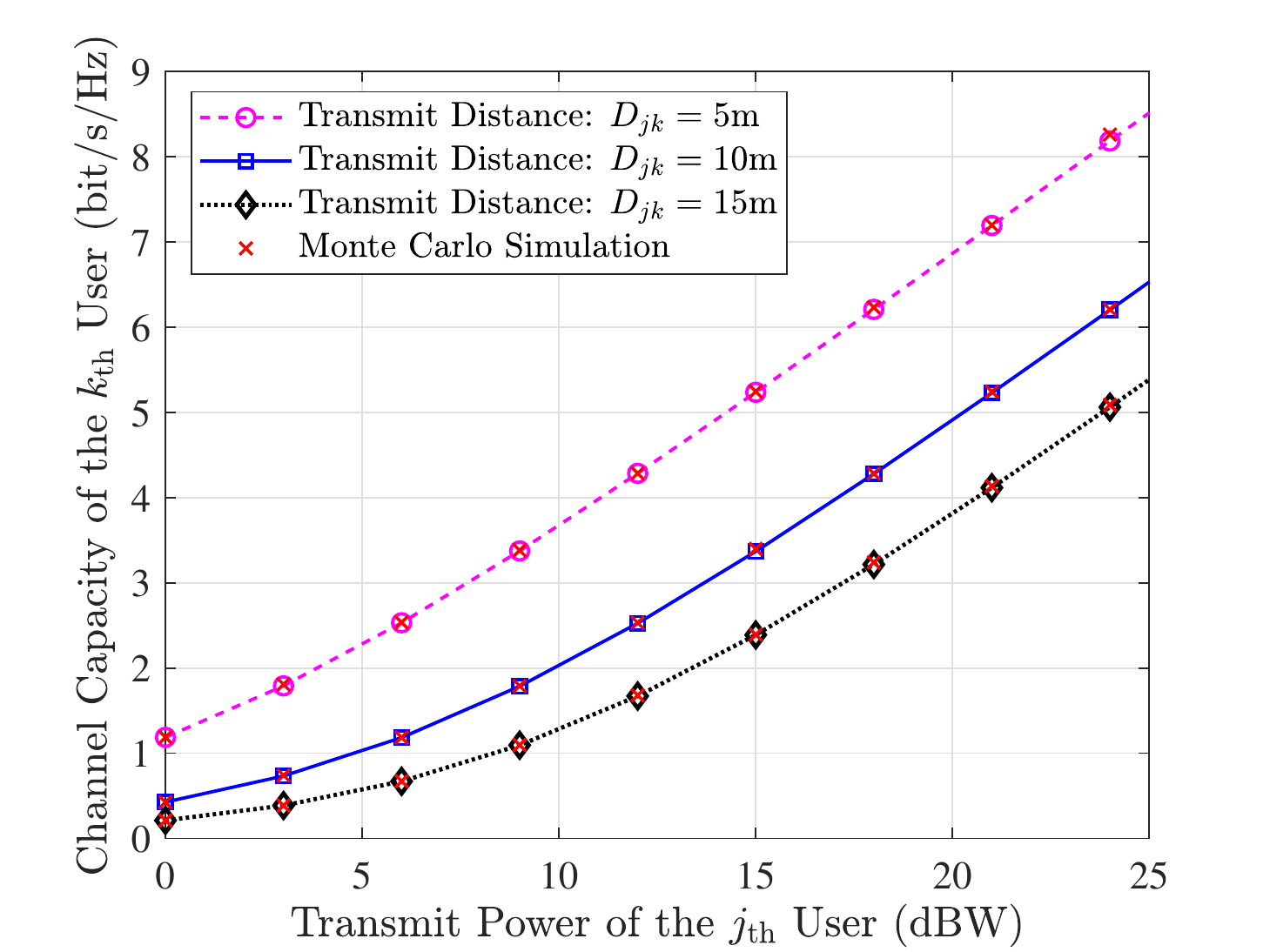}%
			\vspace{0.05cm}
			\caption{The channel capacity of the $k^{\rm th}$ user versus the transmit power of the $j^{\rm th}$ user under different transmit distance, with $D_{jk} = 5$ m, $\beta_{k} = 2$, $\alpha = 4$, $\mu = 5$, $\sigma _N^2 = \sigma _S^2 = 0.1$, $P_j = 10$ ${\rm dBW}$, $P_I = 1$ ${\rm dBW}$, ${\eta _{k}} = 0.2$, ${\upsilon _k}=0.2$, and $N = 2$.}
			\label{CC2}
		\end{minipage}
	\end{figure*}
	
	\subsection{Influence of channel and self-interference}\label{per}
	
	Third, we analyze the impact of wireless channels and self-interference on D2D full-duplex semantic communications. The effect of the ${j^{\rm th}}$ user's (transmitter's) transmit power on the BEP of the ${k^{\rm th}}$ user (receiver) is investigated in Fig.~\ref{BER1}, during which the transmit power of ${k^{\rm th}}$ user is fixed to $20$ dBW. As the transmit power of ${j^{\rm th}}$ user increases from $0$ to $20$ dBW, the BEP of ${k^{\rm th}}$ user decreases from about ${10^{ - 1}}$ to ${10^{ - 5}}$. The clear reason is that since the increase in transmit power makes the SINR higher and thus reduces the BEP. Furthermore, the increase in interfering signal power raises the BEP, and the degree of BEP increase goes up as the transmit power increases. When the transmit power is 0 dBW, for example, the BEPs corresponding to different interfering signal power are relatively close, whereas the BEP differences become larger when the transmit power reaches 20 dBW.

	
	Then, the influence of the ${k^{\rm th}}$ user's transmit power on BEP is analyzed. The experimental results in Fig.~\ref{BER2} demonstrate that, first, with the increase of transmit power, the BEP of ${k^{\rm th}}$ user increases. Taking the case of the channel parameter $\mu{\rm{ = }}2$ as an example, the BEP is smaller than ${10^{ - 4}}$ when the transmit power is 5 dBW, while it reaches nearly ${10^{ - 2}}$ as the transmit power arrives at 30 dBW. This can be interpreted by that the enhancement of transmit power would cause greater self-interference, which further affects the BEP. In addition, when the transmit power is fixed, the BEP decreases as the channel parameter increases, indicating that the severity of multipath effects directly impacts the ${k^{\rm th}}$ user's BEP.
	
	Next, under the condition of different distances between two users, the influence of transmit power on channel capacity is analyzed. From the results in Fig.~\ref{CC2}, one can see that the channel capacity increases with the boost of the transmit power. Taking the case where the distance between the transmitter and receiver is 5 meters as an example, the channel capacity is about 1 bit/s/Hz when the transmit power is 0 dBW, while it reaches 9 bit/s/Hz when the transmit power is increased to 25 dBW. This is reasonable because increasing transmit power raises the SINR. Similarly, increasing the distance between the transmitter and the receiver would lead to a decrease in the SINR. As a result, the larger the distance, the smaller the channel capacity when the transmit power is fixed.
	
	
	\section{Conclusion}\label{S7}
	In this paper, we studied a D2D full-duplex semantic communications system, in which a free-space information sharing mechanism was proposed to avoid perform duplicate computational tasks. Specifically, a SuperPoint-based semantic encoder was presented to extract semantic information, i.e., interest points and descriptors from the view image. Free space information and semantic features are then transmitted to other users in full-duplex communication mode. For the wireless transmission part, we modeled the channel using a generalized small-scale fading model and comprehensively analyzed the effects of self-interference and multiple co-channel signals on the system performance. The close-form channel capacity and BEP were derived. After the receiver receives the semantic information and the free-space computation results of the transmitter, it can perform semantic matching to obtain its own free-space without independent computation.




	\bibliographystyle{IEEEtran}
	\bibliography{Ref}
\end{document}